\documentclass[aps,prd,floats,floatfix,nofootinbib,superscriptaddress,notitlepage]{revtex4-1}
\usepackage{graphicx,amsmath,amssymb,amsfonts,hyperref,mathdots,caption,subcaption, setspace, wrapfig,slashed,units}
\usepackage{tabularx}
\usepackage[usenames,dvipsnames,svgnames,table]{xcolor}
\usepackage[capitalize,nameinlink,noabbrev]{cleveref}

\newcommand{\cah}{\ensuremath{c_{\alpha}}}
\newcommand{\sah}{\ensuremath{s_{\alpha}}}
\newcommand{\cth}{\ensuremath{c_{\theta}}}
\newcommand{\sth}{\ensuremath{s_{\theta}}}
\newcommand{\Tr}[1]{\ensuremath{\,\mathrm{Tr}\left[{#1}\right]}}

\newcommand{\ct}{\ensuremath{^{\dagger}}}
\newcommand{\sw}{s_W}
\newcommand{\cw}{c_W}

\newcommand{\sv}{\ensuremath{\left<\sigma\beta\right>}}

\newcommand{\lamt}[1]{\ensuremath{\widetilde{\lambda}_{#1}}}
\newcolumntype{L}[1]{>{\raggedright\let\newline\\\arraybackslash\hspace{0pt}}m{#1}}
\newcolumntype{C}[1]{>{\centering\let\newline\\\arraybackslash\hspace{0pt}}m{#1}}
\newcolumntype{R}[1]{>{\raggedleft\let\newline\\\arraybackslash\hspace{0pt}}m{#1}}

\begin{document}
\raggedbottom
\allowdisplaybreaks

\title{Dark matter in the Georgi-Machacek model with an additional inert doublet}
\author{Terry Pilkington}
\email{terrydpilkington@gmail.com}
\noaffiliation

\date{\today}

\begin{abstract}
We study a model that extends the Georgi-Machacek model by the addition of an inert doublet, $\Phi_2$. This allows for the model to contain a natural dark matter candidate, $H^0$. For a number of benchmark points at two DM candidate masses ($m_{H^0} \in \{70,\,250\}$~GeV), we determine the relic abundance, direct detection cross section, and collider constraints. We find that the collider studies used do not constrain the model. For the parameter region chosen, the points are ruled out by direct detection experimental results. At the lower mass point, we do not find a viable parameter point, while at the higher mass we do find an additional point that satisfies all the applied constraints. 
\end{abstract}

\maketitle

The discovery of a Standard Model (SM)-like Higgs boson at the Large Hadron Collider (LHC) in 2012~\cite{Aad:2012tfa,Chatrchyan:2012xdj} completed the particle content of the SM. With subsequent runs of the LHC, the Atlas and CMS collaborations have refined measurements~\cite{Khachatryan:2016vau,ATLAS:2016gld,ATLAS:2017cju,ATLAS:2017nkr} of this scalar particle to determine if it is indeed the SM Higgs. Thus far, the measurements agree with predictions of the SM. The remains hope, however, that some room still exists in which Beyond the Standard Model (BSM) physics may hide.

One such BSM model is the Georgi-Machacek (GM) model~\cite{Georgi:1985nv,Chanowitz:1985ug}, which adds two isospin triplets to the original doublet in the Higgs sector. This results in two Higgs particles, two additional neutral scalars, two singly-charged scalars, and one doubly-charged scalar. The GM model may enhance couplings of the Higgs to the $W$ and $Z$ bosons relative to the SM and maintains the relationship between the $W$ and $Z$ boson masses at tree-level, $\rho = \nicefrac{m_W^2}{m_Z^2\,\cos^2\theta_W} = 1$. The phenomenology of the GM model has been very well-studied (see, e.g.,~\cite{Chiang:2015rva,Degrande:2015xnm,Chang:2017niy,Li:2017daq} and references therein).

An important feature of BSM physics models is to include a dark matter (DM) candidate. The Universe consists of a significant amount of DM relative to normal matter such as stars, planets, and manatees. While we have not yet directly observed DM, we do know that it must satisfy certain properties: it must interact with normal matter gravitationally; it does not interact electromagnetically; and it must be stable on cosmological timescales. The Weakly-Interacting Massive Particle (WIMP) paradigm provides a natural DM candidate, one that has the correct relic abundance and that may also interact with the electroweak sector, allowing the possibility of directly detecting DM particles and particle creation at colliders here on Earth. On the other hand, experimental results have not produced a definitive signal of DM~\cite{Arcadi:2017kky,Roszkowski:2017nbc,Hong:2017avi}. The lack of signal allows us to place limits on the interaction cross section between the DM and SM particles.

One of the simplest extensions of the SM is the Inert Doublet Model (IDM)~\cite{Deshpande:1977rw}. The IDM introduces an additional doublet to the SM. This doublet does not acquire a vacuum expectation value (vev) like the usual Higgs doublet. In addition, a $Z_2$ symmetry is imposed on the Lagrangian, making the lightest particle in the inert doublet a DM candidate. The phenomenology of the IDM has been extensively studied (see, e.g., \cite{Branco:2011iw,Blinov:2015vma,Ilnicka:2015ova,Krawczyk:2015xhl,Datta:2016nfz,Hashemi:2016wup,Eiteneuer:2017hoh,Poulose:2016lvz} and references therein).

Unfortunately, none of the additional particles in the GM model are a viable DM candidate. Previously, the GM was furnished with a DM candidate in the form of a singlet scalar~\cite{Campbell:2016zbp}, which examined the effects of requiring that the DM candidate constituted all of the DM density on Higgs coupling measurements. In that study, the authors found that DM candidates in that model were excluded below $57$~GeV, and that Higgs coupling measurements do not constrain the model beyond the constraints on the GM model without the additional singlet scalar. In the current study, we add to the GM model an inert doublet containing a DM candidate, $H^0$, as well as a neutral isoscalar, $A^0$, and a singly-charged scalar, $H^{+}$. We examine the effects of varying the couplings between the triplets and the inert doublet on the relic abundance and direct detection cross section. Additionally, we ensure that the parameter points chosen are not excluded by current LHC searches.

This paper is organized as follows. In \Cref{sec:model}, we introduce the model and fix the notation. In \Cref{sec:param}, we describe the parameter space. We give collider results in \Cref{sec:collider} and DM results in \Cref{sec:dm_constraints}. We conclude in \Cref{sec:conclusions}. Parameter translations are given in appendices.

\section{Model description}
\label{sec:model}
The Georgi-Machacek model extends the SM scalar sector through the addition of two isospin triplets. The usual SM doublet, $\Phi$, carries hypercharge\footnote{We work in the convention $Q = T^3 + \nicefrac{Y}{2}$.} $Y_\Phi = 1$. The complex triplet, $\chi$, carries hypercharge $Y_\chi = 2$ and the real triplet, $\xi$, carries $Y_\xi = 0$. The GM itself does not have a dark matter (DM) candidate, since the neutral particles can decay. To alleviate this shortcoming, we add a second, inert doublet to the GM model.

The Lagrangian describing the combined GM and IDM model is given by
\begin{equation}\label{eq:lagrangian}\begin{aligned}
\mathcal{L} &\supset \left(\mathcal{D}^\mu~\Phi_1\right)\ct\left(\mathcal{D}_\mu~\Phi_1\right) + \left(\mathcal{D}^\mu~\Phi_2\right)\ct\left(\mathcal{D}_\mu~\Phi_2\right) + \left(\mathcal{D}^\mu~\chi\right)\ct\left(\mathcal{D}_\mu~\chi\right) + \frac{1}{2}\left(\mathcal{D}^\mu~\xi\right)\ct\left(\mathcal{D}_\mu~\xi\right) - V\left(\Phi_1,~\Phi_2,~\chi,~\xi\right)\;,
\end{aligned}\end{equation}
where $\mathcal{D}_\mu = \partial_\mu - i \frac{e}{\sw} W^a_\mu T_j^a - i \frac{e}{\cw} B_\mu \frac{Y_j}{2}~$ ($T_j^a$ and $Y_j$ are the generators of $SU(2)_L$ and $U(1)_Y$ respectively where the index $j$ denotes the multiplet) and the multiplet fields are given by
\begin{equation}\label{eq:multiplets}\begin{aligned}
\Phi_{1} &= \left(\begin{matrix}\phi_{1}^{+}\\\phi_{1}^{0}\end{matrix}\right)\;,&
\Phi_{2} &= \left(\begin{matrix}\phi_{2}^{+}\\\phi_{2}^{0}\end{matrix}\right)\;,&
\chi &= \left(\begin{matrix}\chi^{++}\\\chi^{+}\\\chi^{0}\end{matrix}\right)\;,&
\xi &= \left(\begin{matrix}\xi^{+}\\\xi^{0}\\-\xi^{+*}\end{matrix}\right)\;.
\end{aligned}\end{equation}
The vacuum expectation values are given by
\begin{equation}\label{eq:multipletvevs}\begin{aligned}
\left<\Phi_{1}\right> &= \left(\begin{matrix}0\\\frac{v_\phi}{\sqrt{2}}\end{matrix}\right)\;,&
\left<\Phi_{2}\right> &= \left(\begin{matrix}v_{2}^{+}\\\frac{v_{2}}{\sqrt{2}}\end{matrix}\right)\;,&
\left<\chi\right> &= \left(\begin{matrix}0\\0\\v_\chi\end{matrix}\right)\;,&
\left<\xi\right> &= \left(\begin{matrix}0\\ v_\xi \\0\end{matrix}\right)\;.
\end{aligned}\end{equation}
Note that $v_2^{+} \neq 0$ will lead to a charged vacuum, and $v_2 \neq 0$ means that $\Phi_2$ is not inert~\cite{Krawczyk:2015xhl}. Thus, we set $v_2^{+} = v_2 = 0$. To ensure the rho parameter is unity at tree-level (i.e., $m_W^2 = m_Z^2 \cw^2$), we must have $v_\chi = v_\xi$. From the measured value of the Fermi constant, then, we obtain~\cite{Hartling:2014zca}
\begin{equation}\label{eq:fermi_vevs}
v_\phi^2 + 8 v_\chi^2 \equiv v^2 = \frac{1}{\sqrt{2} G_F} \approx \left(246~\unit{GeV}\right)^2\;.
\end{equation}

The scalar potential, $V(\Phi_1,~\Phi_2,~\chi,~\xi) \equiv V_{\mathrm{GM}} + V_{\mathrm{IDM}} + V_{\kappa}$, is given in three parts. To ensure a good DM candidate, we impose a $Z_2$ under which $\Phi_2 \to -\Phi_2$ and all others $x \to x$. In the literature the notation for the scalar quartic couplings in the GM and IDM models overlap. We set here our notation for the combined model. The interactions between the triplets and the non-inert doublet are given by
\begin{equation}\label{eq:VGM}\begin{aligned}
V_\mathrm{GM} &\equiv \frac{\mu_2^2}{2} \Tr{\Phi_{1,M}\ct~\Phi_{1,M}} + \frac{\mu_3^2}{2} \Tr{X\ct~X}\\ &- M_1 \Tr{\Phi_{1,M}\ct~T_2^a~\Phi_{1,M}~T_2^a}\left(U~X~U\ct\right)_{ab} - M_2 \Tr{X\ct~T_3^a~X~T_3^a}\left(U~X~U\ct\right)_{ab}\\ &+ \lamt{1} \Tr{\Phi_{1,M}\ct~\Phi_{1,M}}^2 + \lamt{2} \Tr{\Phi_{1,M}\ct~\Phi_{1,M}} \Tr{X\ct~X} + \lamt{3} \Tr{X\ct~X~X\ct~X}\\ &+ \lamt{4} \Tr{X\ct~X}^2 - \lamt{5} \Tr{\Phi_{1,M}\ct~T_2^a~\Phi_{1,M}~T_2^b} \Tr{X\ct~T_3^a~X~T_3^b}\;,
\end{aligned}\end{equation}
while the interactions between the doublets is given by
\begin{equation}\label{eq:VIDM}\begin{aligned}
V_\mathrm{IDM} &\equiv \mu_{1}^2 \Phi_{2}\ct~\Phi_{2} + \lambda_{2} \left(\Phi_{2}\ct~\Phi_{2}\right)^2 + \lambda_{3} \Phi_{1}\ct~\Phi_{1}~\Phi_{2}\ct~\Phi_{2} + \lambda_{4} \left|\Phi_{1}\ct~\Phi_{2}\right|^2 + \frac{\lambda_{5}}{2} \left[\left(\Phi_{1}\ct~\Phi_{2}\right)^2 + \mathrm{h.c.}\right]\;,
\end{aligned}\end{equation}
and the interactions between the triplets and the inert doublet is given by
\begin{equation}\label{eq:Vkappa}\begin{aligned}
V_{\kappa} &\equiv \kappa_{1}~\Phi_{2}\ct~\Phi_{2}~\chi\ct~\chi + \kappa_{2}~\Phi_{2}\ct~\Phi_{2}~\xi\ct~\xi + \kappa_{3}~\Phi_{2}\ct~T_{2}^{a}~\Phi_{2}~\chi\ct~T_3^a~\chi + \kappa_{4}\left[\widetilde{\Phi}_{2}\ct~T_2^{a}~\Phi_{2}~\chi\ct~T_3^a\xi + \mathrm{h.c.}\right]\;.
\end{aligned}\end{equation}
In \Cref{eq:VGM,eq:VIDM,eq:Vkappa}, $\Phi_{1}$, $\Phi_{2}$, $\chi$, and $\xi$ are as given in \Cref{eq:multiplets}. The bidoublet and bitriplet forms of $\Phi_{1}$, $\chi$, and $\xi$ are
\begin{equation}\begin{aligned}
\Phi_{1,M} &= \left(\begin{matrix} \phi_1^0 & \phi_1^{+} \\ -\phi_1^{+*} & \phi_1^0 \end{matrix}\right)\;,&
X &= \left(\begin{matrix} \chi^{0*} & \xi^{+} & \chi^{++} \\ -\chi^{+*} & \xi^0 & \chi^{+} \\ \chi^{++*} & -\xi^{+*} & \chi^{0} \end{matrix}\right)\;.
\end{aligned}\end{equation}
The matrix, $U$, rotates $X$ into the Cartesian basis is given by
\begin{equation}
U \equiv \frac{1}{\sqrt{2}}\left(\begin{matrix} -1 & 0 & 1 \\ -i & 0 & -i \\ 0 & \sqrt{2} & 0 \end{matrix}\right)\;.
\end{equation}

In the mass basis, the field components are given by
\begin{align}
\phi_{1}^{+} &= \cth G^{+} - \sth H_{3}^{+}\;,\\
\phi_{1}^{0} &= \frac{1}{\sqrt{2}}\left[\cah h + \sah H + v_{\phi} + i \cth G^{0} - i \sth H_{3}^{0}\right]\;,\\
\phi_{2}^{+} &= H^{+}\;,\\
\phi_{2}^{0} &= \frac{1}{\sqrt{2}}\left[H^{0} + i A^{0}\right]\;,\\
\chi^{++} &= H_{5}^{++}\;,\\
\chi^{+} &= \frac{1}{\sqrt{2}}\left[H_{5}^{+} + \cth H_{3}^{+} + \sth G^{+}\right]\;,\\
\chi^{0} &= \frac{\cah H - \sah h}{\sqrt{3}} - \frac{H_{5}^{0}}{\sqrt{6}} + v_{\chi} + \frac{i (\cth H_{3}^{0} + \sth G^{0})}{\sqrt{2}}\;,\\
\xi^{+} &= \frac{\cth H_{3}^{+} - H_{5}^{+} + \sth G^{+}}{\sqrt{2}}\;,\\
\xi^{0} &= \frac{1}{\sqrt{3}}\left[ 2 H_{5}^{0} + \cah H - \sah h\right] + v_{\chi}\;.\\
\end{align}
This results in nine new scalars in addition to the usual Higgs. The $H_3^0$ and $H_3^{\pm}$ are degenerate in mass, as are the $H_5^0$, $H_5^{\pm}$, and $H_5^{\pm\pm}$.
In the above, the mixing angles ($\sah = \sin\alpha_H$, $\cah = \cos\alpha_H$, $\sth = \sin\theta_H$, and $\cth = \cos\theta_H$) are defined by
\begin{equation}\label{eq:mixing_angles}\begin{aligned}
\sin2\alpha_H &\equiv \frac{2 \mathcal{M}_{12}^2}{m_H^2 - m_h^2}\;,&
\cos2\alpha_H &\equiv \frac{\mathcal{M}_{22}^2 - \mathcal{M}_{12}^2}{m_H^2 - m_h^2}\;,\\
\sin\theta_H &\equiv \frac{v_\phi}{v}\;,&
\cos\theta_H &\equiv \frac{\sqrt{8} v_\chi}{v}\;,
\end{aligned}\end{equation}
where
\begin{equation}\begin{aligned}
\mathcal{M}_{11}^2 &= 8\lamt{1} v_\phi^2\;,\\
\mathcal{M}_{12}^2 &= \frac{\sqrt{3}}{2} v_\phi \left[-M_1 + 4 \left(2 \lamt{2} - \lamt{5}\right) v_\chi\right]\;,\\
\mathcal{M}_{22}^2 &= \frac{M_1 v_\phi^2}{4 v_\chi} - 6 M_2 v_\chi + 8 \left(\lamt{3} + 3 \lamt{4}\right) v_\chi^2\;.
\end{aligned}\end{equation}

The physical masses are given in terms of the Lagrangian parameters by
\begin{align}
m_h^2 &= \sah^2 \left[\frac{M_1 v_\phi^2}{4 v_\chi}-6 M_2 v_\chi+8 \left(\lamt{3}+3 \lamt{4}\right) v_\chi^2\right]+\sqrt{3} v_\phi \sah \cah \left[M_1+4 \left(\lamt{5}-2 \lamt{2}\right) v_\chi\right]+8 \lamt{1} v_\phi^2 \cah^2\;,\\
m_H^2 &= \mathcal{M}_{11}^2 + \mathcal{M}_{22}^2 - m_h^2\;,\\
m_3^2 &= \frac{v^2 \left(M_1+2 \lamt{5} v_\chi\right)}{4 v_\chi}\;,\\
m_5^2 &= \frac{v_\phi^2 \left(M_1+6 \lamt{5} v_\chi\right)}{4 v_\chi}+12 M_2 v_\chi+8 \lamt{3} v_\chi^2\;,\\
m_{H^0}^2 &= \frac{1}{2} \left\{2 \mu_1^2+\left[\kappa_3+2 \left(\kappa_1+\kappa_2+\sqrt{2} \kappa_4\right)\right] v_\chi^2+\left(\lambda_3+\lambda_4+\lambda_5\right) v_\phi^2\right\}\;,\\
m_{A^0}^2 &= m_{H^0}^2-2 \sqrt{2} \kappa_4 v_\chi^2-\lambda_5 v_\phi^2\;,\\
m_{H^{+}}^2 &= \frac{1}{2} \left[2 m_{H^0}^2-2 \left(\kappa_3+\sqrt{2} \kappa_4\right) v_\chi^2-\left(\lambda_4+\lambda_5\right) v_\phi^2\right]\;,
\end{align}
where we have eliminated $\mu_1^2$, $\mu_2^2$, and $\mu_3^2$ in favour of $m_{H^0}$ and the vevs. For the purposes of computation, we further eliminate $M_1$, $M_2$, $\lamt{1}$, $\lambda_3$, and $\lambda_4$ in terms of physical masses and vevs. The expression for these translations are given in \Cref{app:mass_trans}. Further, we present the Feynman rules for the model in \Cref{app:feynman}.

\section{Parameter selection}
\label{sec:param}
The GM+IDM model consists of sixteen free parameters. A scan over such a large parameter space would be unfeasible. As such, we must first reduce the number of free parameters. Since we are interested in the model containing a DM candidate, the relic abundance of $H^0$ is an important quantity. Parameters that do not affect the relic abundance can therefore be set to a fixed value.

We first implement the model in \texttt{FeynRules} (v.2.3.29)~\cite{Alloul:2013bka} and calculate the relevant quantities using \texttt{MadGraph5\_aMC@NLO} (v.2.6.0)~\cite{Alwall:2014hca} with the \texttt{MadDM} (v.2.0) plugin~\cite{Backovic:2013dpa,Backovic:2015tpt}. We perform a rough scan of the relic abundance, $\Omega h^2$, and determine the following.
\begin{itemize}
\item{$\lamt{1}$, $\lambda_{4,5}$ are fixed by the masses.}
\item{$\lamt{2}$ has a modest effect on $\Omega h^2$ (up to $\sim20\%$).}
\item{$\lamt{3,4,5}$ have no appreciable effect on $\Omega h^2$ (less than $\sim1\%$).}
\item{$\lambda_{2,3}$ have a minimal effect on $\Omega h^2$ ($\sim5\%$).}
\item{Modifying $\kappa_i$ has the strongest overall effect on $\Omega h^2$ (up to $\sim40\%$).}
\end{itemize}
Our parameter scan will thus
\begin{itemize}
\item{set $m_h = 125$~GeV, $v = (\sqrt{2}G_f)^{-1} \approx 246$~GeV, and $\tan\theta_H = 0.1$;}
\item{set $\lambda_2 = \lambda_3 = \lamt{3} = \lamt{4} = \lamt{5} = 0.1$; and}
\item{scan over $\lamt{2}$, $\kappa_i$, and the masses as given in~\Cref{tbl:param_vals}.}
\end{itemize}

Values were chosen to fall within the allowed ranges based on previous studies of the IDM~\cite{Ilnicka:2015ova} as well as the GM model~\cite{Hartling:2014xma}. This results in 216 parameter points which give a spread of allowed relic abundance values. As we will see later, this set is ruled out by direct detection experimental limits. At the lower DM mass value, $m_{H^0} = 70$~GeV, we do not find parameter values that evade both DM constraints. At the higher DM mass value, $m_{H^0} = 250$~GeV, we find a point that satisfies all of the constraints applied in this study. These points will be further described in~\Cref{sec:dm_direct_detection}.

As shown by the Fermi-LAT collaboration in~\cite{Ackermann:2015zua}, were we to consider indirect detection constraints from dwarf spheroidal galaxies, the Inert Doublet Model can be ruled out by the annihilation to $b\overline{b}$ and $\tau^{+}\tau^{-}$ for DM masses up to $\sim100$~GeV. Similarly, this rules out the parameter points at $m_{H^0} = 70$~GeV through strong couplings $hH^{0}H^{0}$ and $HH^{0}H^{0}$. In~\cite{Queiroz:2015utg}, it is shown that the Cherenkov Telescope Array would be able to rule out DM candidates in the IDM with masses up $\sim800$~GeV. This would additionally rule out our parameter points at $m_{H^0} = 250$~GeV.

In~\cite{Braathen:2017izn,Krauss:2017xpj}, it was shown that loop corrections to the masses $m_3$ and $m_5$ are large in the GM model, leading to a breakdown of perturbativity. This could lead to problems with the higher mass parameter points, and would need to be considered in more detail for this model.

\begin{table}
\begin{center}
{\def\arraystretch{1.35}
\begin{tabular}{C{16em}||C{24em}}
Parameter & Set of values\\
\hline\hline
$\lamt{2}$ & $\left\{0,\,0.1,\,\pi\right\}$\\
$\kappa_1$ & $\left\{0,\,0.1,\,\pi\right\}$\\
$\kappa_2$ & $\left\{0,\,\pi\right\}$\\
$\kappa_3$ & $\left\{0,\,0.1,\,\pi\right\}$\\
$\kappa_4$ & $\left\{0,\,\pi\right\}$\\
$\left(m_{H^0},\,m_{A^0},\,m_{H^{+}},\,m_3,\,m_5\right)$ & $\left\{\left(70,\,110,\,130,\,330,\,350\right),\,\left(250,\,300,\,320,\,830,\,850\right)\right\}$\\
\end{tabular}}
\end{center}
\caption{Parameter values for benchmark points used in this study. Quartic couplings are unitless, while masses are given in $\unit{GeV}$.}
\label{tbl:param_vals}
\end{table}

\section{LHC constraints}
\label{sec:collider}

A model that adds new particles to the SM, when those particles are below the energy reach of current collider experiments, must be compared against the results of those experiments. One of the goals of the Large Hadron Collider (LHC) programme is to find BSM physics. As of yet, there are no definitive signals. The Atlas and CMS collaborations cannot look for every model of BSM physics, and so we must recast their searches to place limits on our models.

One such tool that can take such recast searches is \texttt{CheckMATE 2}~\cite{Dercks:2016npn} which uses \texttt{MadGraph5\_aMC@NLO}~\cite{Alwall:2014hca}, along with \texttt{Pythia 8}~\cite{Sjostrand:2014zea}, \texttt{Delphes 3}~\cite{deFavereau:2013fsa}, and \texttt{FastJet}~\cite{Cacciari:2011ma} to determine if a given parameter point is allowed or excluded by a set of LHC searches. The quantity of interest is $r \equiv \nicefrac{S}{S_{95}}$, where $S$ is the number of signal events computed by \texttt{CheckMATE 2} and $S_{95}$ is the model-independent upper $95\%$ confidence limit from the given search. Values $r <1$ are allowed by that search and $r > 1$ are excluded.

For the GM+IDM model, we use searches from the LHC at two centre-of-mass energies: $8$~TeV~\cite{Chatrchyan:2013mys, Khachatryan:2014rra, Khachatryan:2015lwa, Khachatryan:2015nua, CMS:2013ija, Khachatryan:2014dka, Aad:2013wta, Aad:2013ija, Aad:2014qaa, Aad:2014mha, Aad:2014pda, Aad:2014wea, Aad:2014kra, Aad:2014nra, Aad:2014tda, Aad:2015jqa, Aad:2015zva, Aad:2015wqa, Aad:2015pfx, ATLAS:2012tna, TheATLAScollaboration:2013hha, TheATLAScollaboration:2013tha, TheATLAScollaboration:2013via, ATLAS:2015yda,Aad:2014nua} and $13$~TeV~\cite{Aad:2016tuk, Aaboud:2016uro, Aaboud:2016tnv, Aaboud:2016zdn, Aad:2016qqk, Aad:2016eki, Aaboud:2016lwz, TheATLAScollaboration:2015nxu, TheATLAScollaboration:2016gxs, ATLAS:2016ljb, ATLAS:2016xcm, ATLAS:2016uwq, Aaboud:2017ayj}. The following process is used in \texttt{CheckMATE 2}. First, in \texttt{MadGraph5\_aMC@NLO}, define $new$ as the set of all particles added to the SM by GM+IDM and then produce $50000$ $p p \to new\,new$ events at each parameter point at each selected centre-of-mass energy. These events are passed to \texttt{Pythia 8} to shower, and subsequently \texttt{Delphes 3} for detector simulation. \texttt{CheckMATE 2} then uses an \texttt{AnalysisHandler} to prepare the reconstructed events for analysis, which returns the number of events that satisfy criteria for each signal region in the given analysis. The scan points chosen all pass the requirements. The highest $r$ values obtained in each energy regime for this study are shown in~\Cref{tbl:cm_r_values}.

For the $8$~TeV point, the strongest constraint comes from a search for direct production of charginos and neutralinos in events with three leptons and missing energy~\cite{Aad:2014nua}. The signal region that gives this (not very strong) constraint, \texttt{SR0$\tau$a02}, requires two same flavour, opposite sign (SFOS) leptons with invariant mass $12-40$~GeV and $\slashed{E}_T > 90$~GeV. The largest cross section in our model at this point comes from $p\,p \to H^0\,H^{\pm}$, which can decay thorough $H^0\,A^0\,W^{\pm*} \to H^0\,H^0\,Z^{*}\,W^{\pm*} \to H^0\,H^0\,\ell^{+}\,\ell^{-}\,\ell'\,\nu_{\ell'}$. Events such as this will indeed contain a large amount of $\slashed{E}_T$, since they involve two DM particles and a neutrino. The SFOS leptons will fall within the required invariant mass range because the maximum energy available to the off-shell $Z$ boson will be $40$~GeV.

For the $13$~TeV point, the strongest constraint comes from a search for scalar top partner pair production in events with four or more jets and missing energy~\cite{Aaboud:2017ayj}. The signal region that gives this (not very strong) constraint, \texttt{SRA-T0}, requires two $b$ jets and $\slashed{E}_T > 550$~GeV. The largest cross section in our model at this point comes from $p\,p \to H^0\,H^{\pm}$, which can decay thorough $H^0\,A^0\,W^{\pm*} \to H^0\,H^0\,Z^{*}\,W^{\pm*} \to H^0\,H^0\,b\,\overline{b}\,\ell'\,\nu_{\ell'}$. For this process to met the $\slashed{E}_T$ requirement, the $H^0$ must be highly boosted.

\begin{table}
\begin{center}
{\def\arraystretch{1.35}
\begin{tabular}{C{5em}C{3em}C{3em}C{3em}C{3em}C{3em}C{3em}||C{7em}C{6em}}
Energy & $m_{H^0}$ & $\lamt{2}$ & $\kappa_1$ & $\kappa_2$ & $\kappa_3$ & $\kappa_4$ & $r$ & LHC search\\
\hline\hline
$8$~TeV & $70$ & $0.1$ & $0.1$ & $0$ & $\pi$ & $\pi$ & $2.51 \times 10^{-4}$ & \cite{Aad:2014nua}\\
$13$~TeV & $70$ & $0.1$ & $\pi$ & $\pi$ & $\pi$ & $\pi$ & $1.33 \times 10^{-3}$ & \cite{Aaboud:2017ayj}\\
\end{tabular}}
\end{center}
\caption{Parameter points in the scan region with the largest $r = \nicefrac{S}{S_{95}}$ value. A value of $r>1$ would be excluded by the given search.}
\label{tbl:cm_r_values}
\end{table}

\section{Dark matter constraints}
\label{sec:dm_constraints}

\subsection{Relic abundance}
\label{sec:dm_relic_abundance}

The relic abundance of a particle species---how much of it exists in the Universe---is an important quantity.The Planck collaboration finds~\cite{Ade:2015xua}
\begin{equation}
\Omega_{\rm DM} h^2 = 0.1188 \pm 0.0010\;.
\end{equation}
Given a standard freeze-out, the number density of a particle species, $n$, follows~\cite{Kolb:1990vq}
\begin{equation}\label{eq:numb_density}
\frac{dn}{dt} + 3 H n = -2 \sv \left(n^2 - n_{\rm eq}^2\right)\;,
\end{equation}
where $H$ is the Hubble constant, $\sv$ is the thermally-averaged annihilation cross section, and $n_{\rm eq}$ is the number density of the particle in thermal equilibrium with the Universe at a given temperature. The relic abundance is the proportional to the current number density.

\begin{figure}
\centering
	\begin{subfigure}{0.4\textwidth}
		\includegraphics[width=\textwidth]{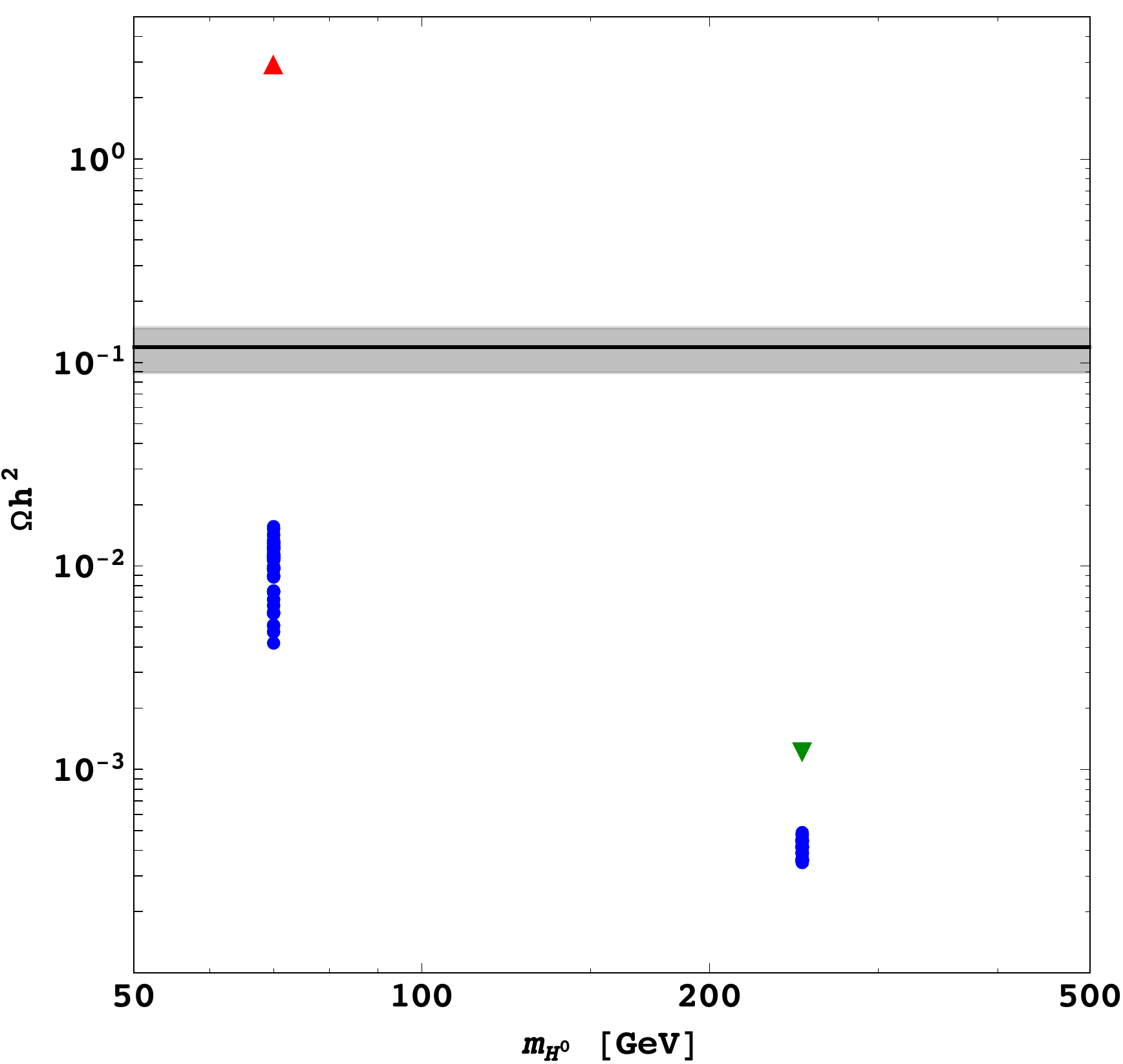}
		\caption{}
		\label{fig:GMIDM_Omegah}
	\end{subfigure}
	\hspace*{1em}
	\begin{subfigure}{0.4\textwidth}
		\includegraphics[width=\textwidth]{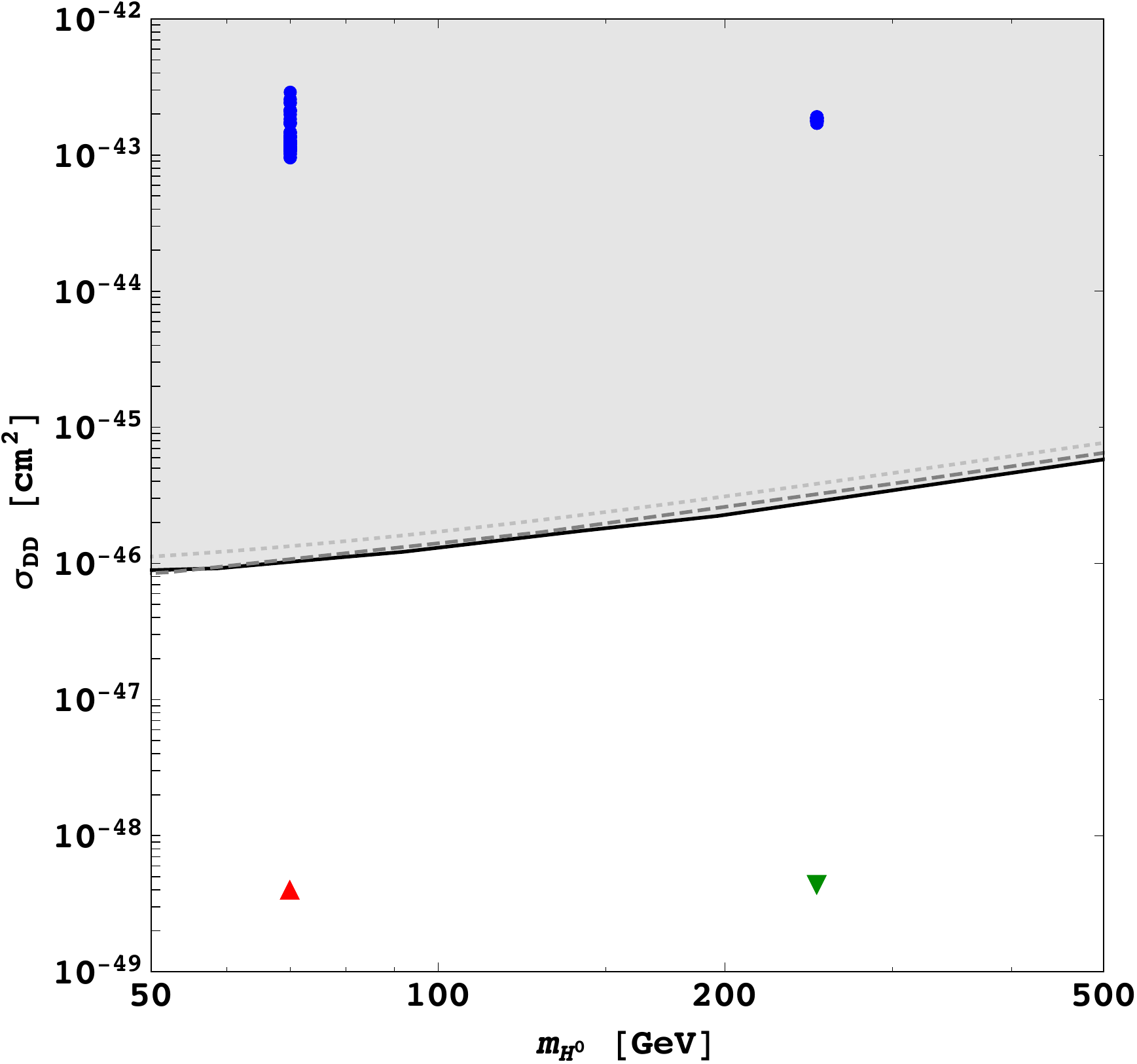}
		\caption{}
		\label{fig:GMIDM_sigmaDD}
	\end{subfigure}
	\caption{Left panel shows relic abundance as a function of DM candidate ($H^0$) mass. The solid black line is the relic abundance from the Planck experiment~\cite{Ade:2015xua}, with the grey region representing a $25\%$ uncertainty. Right panel shows direct detection cross section as a function of DM candidate ($H^0$) mass. The solid black line show the bound from the PandaX-II~\cite{Cui:2017nnn} experiment, the dashed medium grey line shows the Xenon1T~\cite{Aprile:2017iyp} bound, and the dotted light grey line shows the LUX~\cite{Akerib:2016vxi} bound. In both plots, the blue circular points are the scan points chosen in~\Cref{sec:param}, while the triangular points are the parameter points chosen to evade direct detection bounds, as described in~\Cref{sec:dm_direct_detection}.}
	\label{fig:GMIDM_plots}
\end{figure}

We compute the relic abundance for our DM candidate, $H^0$, using \texttt{MadDM}~(v.2.1.0)~\cite{Backovic:2013dpa}, a plug-in for \texttt{MadGraph5\_aMC@NLO}~(v. 2.6.0)~\cite{Alwall:2014hca}. \texttt{MadDM} determines the relic abundance using all $2 \to 2$ processes with the DM candidate in the initial states, and numerically solves~\Cref{eq:numb_density}. For the points described in~\Cref{sec:param}, the resulting relic abundance is shown as the blue circles in~\Cref{fig:GMIDM_Omegah}. From this, we see that coupling between the inert doublet and the triplets does indeed affect the relic abundance. At $m_{H^0} = 70$~GeV, varying $\lamt{2}$ and $\kappa_i$ causes a factor of 4 variation in the relic abundance about $\Omega h^2 \approx 0.01$, while at $m_{H^0} = 250$~GeV we find a factor of 2 variation in the relic abundance about $\Omega h^2 \approx 0.0004$. Note that all of the values fall well below the measured value of $\Omega h^2 = 0.1188$, shown as a solid black line with an uncertainty band of $25\%$ in grey on the plot, meaning that if $H^0$ is the DM at these parameter points, it cannot be the only component of the DM.

\subsection{Direct detection}
\label{sec:dm_direct_detection}

If a DM candidate interacts non-gravitationally with SM particles, then we should be able to detect it here on Earth. To look for Weakly-Interacting Massive Particles (WIMPs) such as our $H^0$, experiments use large volumes of liquid noble gas (like xenon or argon) in low-background environments and attempt to detect the small deposit of energy left by a WIMP colliding with a nucleus. Thus far, there has been no definitive signal from a DM-nucleon collision in experiments. The lack of signal allows us to place limits on the interaction cross section, with the most sensitive limits currently from LUX~\cite{Akerib:2016vxi}, PandaX-II~\cite{Cui:2017nnn}, and Xenon1T~\cite{Aprile:2017iyp}.

The direct detection cross section is computed also using \texttt{MadDM}~(v.2.1.0)~\cite{Backovic:2013dpa}, which numerically computes the low-energy effective operator coefficients. Using these coefficients, \texttt{MadDM} calculates the relevant $2 \to 2$ scattering processes for DM interacting with a nucleon. The resulting cross sections for our parameter scan are shown as the blue circles in~\Cref{fig:GMIDM_sigmaDD}. In this figure, we also plot the limit from LUX~\cite{Akerib:2016vxi} (dotted light grey), PandaX-II~\cite{Cui:2017nnn} (solid black), and Xenon1T~\cite{Aprile:2017iyp} (dashed medium grey). The shaded grey region above the curves represents the parameter space excluded by the experimental results. As we see, the blue circles from our scans lie entirely in the excluded region.

\begin{table}
\begin{center}
{\def\arraystretch{1.35}
\begin{tabular}{C{4em}C{4em}C{4em}C{4em}C{4em}C{4em}||C{9em}C{4em}}
$m_{H^0}$ & $m_{A^0}$ & $m_{H^{+}}$ & $m_{3}$ & $m_{5}$ & $\lambda_{3}$ & $\sigma_{\rm DD}$ & $\Omega h^2$\\
\hline\hline
70 & 110 & 130 & 330 & 350 & $0.398$ & $4.15\times 10^{-49}$~cm$^2$ & $2.99$\\
250 & 300 & 320 & 830 & 850 & $1.33$ & $4.52\times 10^{-49}$~cm$^2$ & $0.0012$
\end{tabular}}
\end{center}
\caption{Parameter values chosen to evade direct detection bounds, with their resulting relic abundance and direct detection cross section. Quartic couplings are unitless, while masses are given in $\unit{GeV}$. $\lamt{2,3,4,5} = 0.1$ $\kappa_i = 0.1$ $\lambda_{2} = 0.1$ $\lambda_{4,5}$ are defined by the masses.}
\label{tbl:fine_tuned}
\end{table}

We would like to find some point that evades the direct detection limit. We use the values given in~\Cref{tbl:fine_tuned}. At $m_{H^0} = 70$~GeV, we find that, by moving into the part of parameter space that allows the direct detection limit to be satisfied, we break the limit on the relic abundance with $\Omega h^2 = 2.99$. This point is shown as the red up triangle in~\Cref{fig:GMIDM_plots}. This is caused by the coupling of $H^0$ to $h$, which has a singularity when the mass difference between the two Higgs particles is small, $|m_{H} - m_{h}| \approx 0$. The coupling is given by
\begin{equation}
g_{h\,H^0\,H^0} = \frac{1}{3} i \left\{\sqrt{3} \left[\kappa_3+2 \left(\kappa_1+\kappa_2+\sqrt{2} \kappa_4\right)\right] s_\alpha v_\chi - 3 \left(\lambda_3+\lambda_4+\lambda_5\right) c_\alpha v_\phi\right\}\;.
\end{equation}
However, at $m_{H^0} = 250$~GeV, we find that the point is allowed, with relic abundance $\Omega h^2 = 0.0012$ and direct detection cross section $\sigma_{\rm dd} = 4.52\times 10^{-49}$~cm$^2$. The singularity at $\alpha_H \approx 0$ does not occur at this parameter point because the mass difference between the two Higgs particles is much greater. Both of these points are allowed by the collider constraints of~\Cref{sec:collider}.

\section{Summary and outlook}
\label{sec:conclusions}

In this paper, we studied the Georgi-Machacek model extended by an inert doublet in order to furnish the GM model with a dark matter candidate. Our interest was in how the couplings between the inert doublet and the triplets affected the collider prospects, relic abundance, and direct detection limits. Over a selection of points in the $(m_{H^0},\,\lamt{2},\,\kappa_1,\,\kappa_2,\,\kappa_3,\,\kappa_4)$ plane, we find that collider and relic abundance constraints are satisfied, but the direct detection cross sections fall in the region excluded by current experimental limits.

We then looked for a point to evade the direct detection limits at the two mass values used. At the lower mass, $m_{H^0} = 70$~GeV, we find that evading the direct detection limit causes the relic abundance to be too large. On the other hand, at the higher mass, $m_{H^0} = 250$~GeV, we find a point that satisfies both direct detection and relic abundance limits.

A full scan over all sixteen free parameters would be the ideal next step, however the computational resources required for this would be great. Further, constraints arising from indirect dark matter experiments should be applied to the GM+IDM model. A more detailed study of the collider constraints on the model near the higher mass scale would also be intriguing.

\begin{acknowledgments}
This work was supported by Harold \& Frances Tarris, Michael \& Pauline Pilkington, Jarrett \& C\'{e}line Derksen, David \& Dilys Bowes, and Clarke \& Rhian Piprell.

The author would like to thank Simon Fraser University and the University of the Fraser Valley for access to their libraries. We would like to thank Katy Hartling, Heather Logan, and Roozbeh Yazdi for discussion and comments. We would also like to thank Jamie Tattersall for technical assistance with \texttt{CheckMATE 2} and Olivier Mattelaer for technical assistance with \texttt{MadGraph5\_aMC@NLO} and \texttt{MadDM}.

The author would like to acknowledge the St\'{o}:l\={o}, Tsleil Waututh, Katzie, Musqueam, and Squamish peoples on whose traditional territories this work was done.
\end{acknowledgments}

\appendix

\section{Transformations}
\label{app:mass_trans}
To make computation easier, we transform certain Lagrangian parameters in terms of masses. These transformations are listed below.
\begin{align}
\lamt{1} &= \frac{1}{8}\left(m_h^2 + \frac{\mathcal{M}_{12}^2}{\mathcal{M}_{22}^2 - m_h^2}\right)\;,\\
\mu_1^2 &= m_{H^{+}}^2 - \frac{1}{2}\left[2(\kappa_1 + \kappa_2) - \kappa_3\right] v_\chi^2 - \frac{1}{2} \lambda_3 v_\phi^2\;,\\
\mu_2^2 &= -6\left(\lamt{2} - \frac{m_3^2}{v^2}\right) - 4 \lamt{1} v_\phi^2\;,\\
\mu_3^2 &= \frac{m_5^2}{2} - 4\left(2 \lamt{3} + 3 \lamt{4}\right) v_\chi^2 - 2\left(\lamt{2} - \frac{m_3^2}{4 v^2}\right) v_\phi^2\;,\\
M_1 &= 2 \left(\frac{2 m_3^2}{v^2} - \lamt{5}\right) v_\chi\;,\\
M_2 &= \frac{1}{12 v_\chi}\left[m_5^2 - \left(\lamt{5} + \frac{m_3^2}{v^2}\right) v_\phi^2\right]\;,\\
\lambda_4 &= \frac{1}{v_\phi^2}\left(m_{H^0}^2 + m_{A^0}^2 - 2 m_{H^{+}}^2 - 2 \kappa_3 v_\chi^2\right)\;,\\
\lambda_5 &= \frac{1}{v_\phi^2}\left(m_{H^0}^2 - m_{A^0}^2 - \sqrt{8} \kappa_4 v_\chi^2\right)\;.
\end{align}


\end{document}